\documentclass[%
 reprint,
 amsmath,amssymb,
 aps,
floatfix
]{revtex4-2}

\def\rmdj {\rlap{\kern 0.25em\raise 1.2ex\hbox
  {\vrule height 0.10ex width 0.28em}}d}
\def\sldj {\rlap{\kern 0.25em\raise 1.3ex\hbox
  {\vrule height 0.10ex width 0.28em}}d}
\def\itdj {\rlap{\kern 0.25em\raise 1.3ex\hbox
  {\vrule height 0.10ex width 0.28em}}d}
\def\bfdj {\rlap{\kern 0.25em\raise 1.3ex\hbox
  {\vrule height 0.10ex width 0.28em}}d}
\def\ttdj {\rlap{\kern 0.25em\raise 1.3ex\hbox
  {\vrule height 0.10ex width 0.28em}}d}
\def\scdj {\rlap{\kern 0.25em\raise 1.3ex\hbox
  {\vrule height 0.10ex width 0.28em}}d}
\def\sfdj {\rlap{\kern 0.25em\raise 1.3ex\hbox
  {\vrule height 0.10ex width 0.28em}}d}

\def\dj{\ifvmode\leavevmode\fi\ifcase\fam \rmdj \or \or \or
  \or \itdj \or \sldj \or \bfdj \or \ttdj \or \sfdj \or \scdj \else \rmdj \fi}

\usepackage{tikz}
\usetikzlibrary{arrows,matrix}
\usepackage{relsize}

\newcommand{\Emmett}[5]{
\draw[#4] (0,0)
\foreach \x in {1,...,#1}
{   -- ++(rand*#2,rand*#3)
}
node[right] {#5};
}

\def\ups{{\scriptstyle \Upsilon}_{_0}}

\def\f{\frac}
\def\p{\partial}
\def\be{\begin{equation}}
\def\ee{\end{equation}}

\usepackage{graphicx}
\usepackage{dcolumn}
\usepackage{bm}

\begin{document}

\preprint{APS/123-QED}

\title{On the Extension of a Physical Body in Classical Motion.
\\
An Analogy for a Pseudo-Velocity Concept and Wiener's Process \\ in (Ideal) Polymer Solutions}

\author{Stefano A. Mezzasalma}
\affiliation{Materials Physics Division, Ru\dj er Bo\v{s}kovi\'c Institute, Bijeni\v{c}ka cesta 54, 10000 Zagreb, Croatia.
\\
Lund Institute for advanced Neutron and X-ray Science (LINXS), University of Lund, Ideon Building: Delta 5, Scheelev\"{a}gen 19, 223 70 Sweden.\\
E-mail: smezzas@irb.hr}


\begin{abstract}

A pseudo-velocity concept, based on the extension of a linear body, is defined by a special relativity experiment.
It suggests an analogy with the covariance properties of Wiener's process, ultimately implying that the scaling behavior of (Gaussian) polymer solutions can be derived from a Brownian Relativity theory, as it was formerly put forward.
An ad-hoc statistical interpretation of the resulting spacetime transforms may be given by the central limit theorem.

\end{abstract}

\maketitle


\section*{Introduction}

\noindent Brownian relativity (BwR) is the name given to a framework that combines the conceptual structures of relativity and Brownian motion \cite{sam1}.
It propounds that time and space in a Brownian system can be envisaged similar to the spacetime of Einstein’s relativity.
To avoid misinterpretations, it is worth perhaps clarifying from the outset that BwR does not concern or want to contribute to long-standing problems of relativistic Brownian motion and heat transport, which deal with issues as covariance and causality violations by the propagator of non-relativistic diffusion, the (non-)existence of Markov's processes and random space-like curves that are covariant upon the (inhomogeneous) Lorentz group \cite{dud66,hak68,dun09}.
BwR was rather restricted to studying (universal limits in) polymer solutions, building up a language and analogies that would only conflict with the previous points if BwR were extended and applied outside its formal and phenomenological domains.\\
Accordingly, average size and characteristic time of a polymer fluctuating in a liquid were derived similarly to a Lorentz–FitzGerald length contraction and a time dilation rule, if the system is short-range correlated (or uncorrelated), and to an equivalence relation for geometry and statistics when correlations are long-ranged.
BwR predicts new universal laws for the relevant scaling exponents in polymer solutions, i.e. for chain size ($\nu$), diffusion coefficient ($\delta$), characteristic time ($\sigma$) and viscosity ($\varepsilon$), which are fulfilled by theoretical and experimental values in different molecular weight and volume fraction regimes, in both unentangled and entangled systems (see e.g. reptation \cite{pierre79} and renormalization \cite{nie82} theories, athermal, good and $\Theta$ solvents \cite{rub03}), e.g. \cite{sam1,sam2}:
\be
\delta + \nu \; = \; 3 \nu - \sigma \; = \; - \; \tfrac{\varepsilon}{2} 
\label{eqn:1}
\ee
Such an agreement was seemingly not inquired in the current literature, nor the link BwR settles, between molecular and macromolecular scales \cite{sam4,sam5}, was tested in disciplines of utmost importance such as turbulence in liquids, where the idea to probe the statistical properties of a turbulent flow by large molecules (whether they are called fibers, polymers, or material lines) has attracted attention quite recently \cite{bro14,ver16,theo17}, or life sciences, in which BwR may be promising e.g. to the systems biologist's work of modelling cell pathways at any relevant spatiotemporal scale \cite{nov15}, and especially in the less developed field of intercellular processes or when diffusion is slower than biochemical reactions \cite{cow12}.
Diffusing molecules must get to specific absorbing sub-domains of cell membranes to trigger a reaction, this being normally described by diffusion-reaction models and their master equations \cite{gil07,isaac09}.\\
Purpose of this letter is clarifying and delving into some of the main aspects of the special version of BwR.
To this end, the next section describes a special relativity experiment representative of BwR, where the pseudo-velocity $=$ (linear body extension)/time is introduced in analogy to the role played in BwR by a polymer molecule.

\section*{Observer's Velocity from a Pseudo-Velocity}

\noindent In BwR the perturbation to the local diffusion coefficient induced by a polymer chain in a liquid may be formulated to give rise to a transformation conceptually similar to the Lorentz-Poincar\'e type.
The light speed ($c$) in Special Relativity (SR) is formally replaced here by the diffusion coefficient of liquid molecules ($D_{_1}$), taken to be a constant and homogeneous diffusivity unit throughout the solution.
That, clearly, was only an analogy, as the following SR example is going to expound.
It had nothing to do with either the impassible limit introduced by the second SR postulate, or the related causality prescriptions.\\
A pointwise observable and a rectilinear line of length $\ell_{_0}$ move uniformly between points $A$ and $B$, with $\overline{AB} = \Delta l$.
Let an extreme of the line to lie on $A$, while the other falls within $\overline{AB}$.
From a frame ${\rm K}_{_0}$ at rest with $A$ and $B$, the point and the line reach the extreme $B$ at equal times ($\Delta t$), with velocities $v = \Delta l / \Delta t$ and $v' = (\Delta l - \ell_{_0}) / \Delta t$.
From another frame ${\rm K}_u$, moving with velocity $u$ relative to ${\rm K}_{_0}$, the Lorentz-Poincar\'e transform \cite{mis17} returns two time intervals, for the point and line, equal respectively to:
\be
\Delta t' \; = \; \tfrac{\gamma_u}{v} (1 - \beta_u \beta_v) \Delta l
\ee
\be
\overline{\Delta t'} \; = \; \tfrac{\gamma_u}{v'} (1 - \beta_u \beta_v) (\Delta l - \ell_{_0})
\ee
where $\gamma_w$ generally denotes a Lorentz factor evaluated at a velocity $w$, and $\beta_w = w/c$.
Imposing that ${\rm K}_u$ agrees with the simultaneous arrival in $B$ of the point and line ($\overline{\Delta t'} = \Delta t'$), and let $\ups \equiv \ell_{_0} / \Delta t$ be a pseudo-velocity (in the same direction and orientation of $v$), it turns out:
\be
v' \; = \; \f{v - \ups}{1 - \beta_u \beta_{_0}}
\label{eqn:4}
\ee
still with $\beta_{_0} = \ups / c$.
The velocity $v'$ may clearly belong to a second observer, which travels in the smaller space reduced by $\ell_{_0}$, between the extreme of the line closest to $B$ and point $B$.
The quantity $\ups$ has the dimensions of a velocity as well but, physically, it is {\em not}, and by applying to $\ell_{_0}$ the length contraction rule and the Lorentz-Poincar\'e transformation of times, it transforms as:
\be
\ups' \; = \; \f{1 - \beta^2_v}{1 - \beta_v \beta_{_0}} \ups
\label{eqn:5}
\ee
However, if one regards it as the speed detected from ${\rm K}_{_0}$ of a movement on the spot, travelling along the rectilinear rod, then, for a frame ${\rm K}_v$ linked to the pointwise observer, Eq. (\ref{eqn:4}) yields the expected velocity composition law of SR, i.e.:
\be
v' = \; v \ominus \ups
\label{eqn:6}
\ee
On this basis, we may form now a new quantity, sum of measurements in Eq. (\ref{eqn:5}) and Eq. (\ref{eqn:6}), and observe that it points out an identity, holding $\forall v \le c$:
\be
v \ominus \ups \; = \; v - \ups'
\label{eqn:7-x}
\ee
or, identically:
\be
(v \ominus \ups) \; + \; \ups' \; = \; v
\label{eqn:7}
\ee
the light speed being the only value for which $\ups' = 0$.\\
In conclusion, the observer's $v$ is regained in Eq. (\ref{eqn:7}) by summing the contribution of the motion {\em in place} to the velocity measurement (see the sketch in Fig. 1).

\begin{figure}[b]
\includegraphics[width=\columnwidth, height=5.5cm]{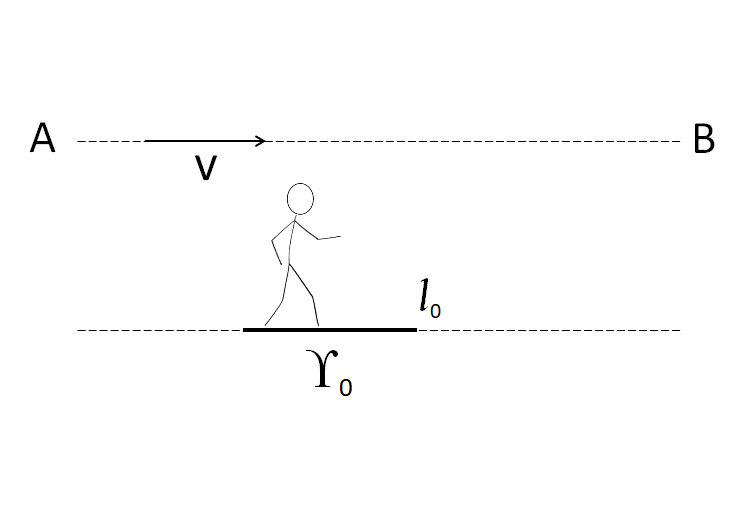} 
\caption{\label{fig:1} Scheme of the experiment bringing to Eq. (\ref{eqn:7}).}
\end{figure}

\section*{Wiener's Process and Brownian Relativity}

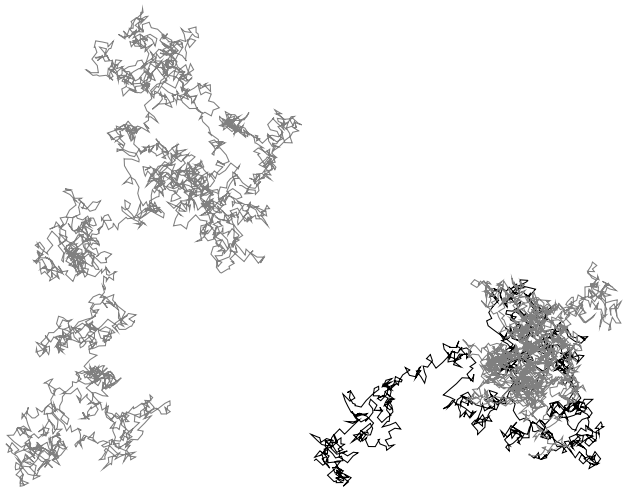
\begin{figure}
\begin{tikzpicture}
\Emmett{3000}{0.1}{0.1}{gray}{}
\end{tikzpicture}
\begin{tikzpicture}
\Emmett{1200}{0.1}{0.1}{black}{}
\Emmett{1800}{0.1}{0.1}{gray}{}
\end{tikzpicture}
\caption{\label{fig:2} Schematic BwR representation ($3000$ steps).
A Brownian liquid molecule fluctuates with diffusion coefficient $D_{_1}$ until time $t$ (grey path).
A chain molecule (black snapshot) substitutes a portion of this path $(0 < t' < t)$ at a constant diffusivity.}
\end{figure}

\noindent The quantity $\ups$ is reminiscent of a self-correlation term for the slower movement, pointing out the sought BwR analogy to SR.
In what follows, neither It\^{o}'s calculus \cite{ito51} nor a phase space representation with momentum coordinate will become necessary.
All should be recalled is a fundamental feature of Wiener-Levy's process ($W (t)$), representing a Brownian motion with time increments much larger than Rayleigh's damping times
\cite{*[{It is known, see Thm. 9.3 in }] [{, Einstein's is a good approximation to Ornstein-Uhlenbeck's theory for a free particle.}] nel67}.
Without losing generalities, let a one-dimensional motion and two time instants, $t > t' \ge 0$.
Since \cite{van07}:
\be
\overline{W (t) W (t')} \; = \; {\rm min} \{ t, t' \}  \; = \; {\rm min} \{ \overline{W^2 (t)}, \overline{W^2 (t')} \}
\label{eqn:8}
\ee
it turns out a relation having the structure of Eq. (\ref{eqn:7}):
\be
\overline{(W (t) - W (t'))^2} \; + \; \overline{W (t) W (t')} \; = \; {\rm max} \{ t,t' \} \; = \; t
\label{eqn:9}
\ee
with the first term accounting for the relative Brownian motion and the second for the self-correlation.
It can be applied to a molecule long $n$ units ($W_n (t')$) and a liquid molecule ($W_{_1} (t)$) sharing the same paths:
\be
\overline{W^2_{_1} (t)} \; + \; \overline{W^2_n (t')} \; - \; \overline{W_{_1} (t) W_n (t')} \; = \; D_{_1} t
\label{eqn:10}
\ee
because, let $p_i (s, x, y) = {\rm e}^{- \f{(x - y)^2}{2 D_i s}} / \sqrt{2 \pi D_i s}$ be the Gaussian transition density, one obtains \cite{*[{When $y=0$ such a density satisfies $\p p_i / \p t = \f{1}{2} \p^2 p_i / \p x^2$, see e.g. }] [{, the space dimensionality factor ($2 d$) in Einstein's law, $\overline{\bf r^2} = 2d D t$, being irrelevant to the final description (Eq. \ref{eqn:20}).
According to the analogy $n \rightarrow t$, for molecular weight and time, the same argument applies to $p_i (n,x,0)$ as well.}] brz00}:
\be
\int_{\Re^2} x y p_n (t', 0, x) p_{_1} (t - t', x, y) d x d y \; = \; D_{n} t' \;\;\; (t > t')
\label{eqn:11}
\ee
Eq. (\ref{eqn:9}) and Eq. (\ref{eqn:10}) express how in BwR the time for a liquid molecule to run across two (radial) positions is regarded as invariant, regardless of chain paths in between.
Eq. (\ref{eqn:9}) may be also translated in random walk terms by the total step number $N$, step size $l$ and Kuhn's length $l'$, e.g.:
\be
\f{(N - n) \; l^2}{D_{_1}} \; + \; \f{l'^2}{D_{n}} \; = \; \f{N l^2}{D_{_1}}
\ee
that returns an identity for a Rouse ideal coil with $l' = l$, $D_n = D_{_1} / n$ accounting for the motion of its gravity center.
These natural assumptions will be preserved below, upon the hydrodynamic limit of large $t$ and the (real) continuation $N, n \rightarrow t, t'$ of polymer physics \cite{pierre79}.\\
Sum of second and third addenda on the left side of Eq. (\ref{eqn:10}) is the zero mean of two uncorrelated increments, i.e. $(W_n (t') - W_{_1} (0)) (W_n (t') - W_{_1} (t))$.
One can perturb it by a random disturbance, and the two sampling frames of reference ${\rm K}_{_1}$, ${\rm K}_{n}$ will measure:
\be
{\rm K}_{_1} (\tau, W^*) : \;\;\;
D_{_1} (t - \Delta^{\alpha} \tau) \; + \; \overline{W_{_1} \Delta^{\alpha} W^*}
\label{eqn:13}
\ee
\be
{\rm K}_{n} (\tau', W'^*) : \;\;\;
D_{_1} (t - \Delta^\alpha \tau') \; + \; \overline{W_n \Delta^\alpha W'^*}
\label{eqn:14}
\ee
the perturbation order being kept unspecified, to be discussed later.
Eq. (\ref{eqn:13}) and Eq. (\ref{eqn:14}) cannot be equated, as $\Delta^\alpha \tau \neq \Delta^\alpha \tau'$ and Eq. (\ref{eqn:10}) prescribe $t$ to be fixed.
This is possible in the limit of large time intervals, when the displacement between dynamic and static variances in ${\rm K}^*_{n}$:
\be
\Delta^\alpha \sigma^2 \; (\tau', W'^*; n) \; = \; D_{_1} \Delta^\alpha \tau' \; - \; \overline{W_n \Delta^\alpha W'^*}
\ee
tends to the value taken on in ${\rm K}^*_{_1}$:
\be
\Delta^\alpha \sigma^2 \; (\tau, W^*; 1) \; = \; \Delta^\alpha \sigma'^{2} \; (\tau', W'^*; n) \;.
\label{eqn:16}
\ee
${\rm K}^*_{n,1}$ indicate the two frames at the hydrodynamic limit, at which the polymer scaling laws were formerly derived from BwR.
They comprise a time, linked to dynamic fluctuations of a Brownian particle, and a measure of space which stems from shape fluctuations (see Eq. \ref{eqn:8}):
\be
\overline{W_{_1} \Delta^\alpha W^*} \; = \; \overline{(\Delta^\alpha W^*)^2} \; \equiv \; \Delta^\alpha \overline{\delta^2_{_1}}
\label{eqn:17}
\ee
\be
\overline{W_n \Delta^\alpha W'^*} \; = \; \overline{(\Delta^\alpha W'^*)^2} \; \equiv \; \Delta^\alpha \overline{\delta^2_n}
\label{eqn:18}
\ee
The relative diffusivity is pointed out as usual by the first average on the left side of Eq. (\ref{eqn:9}), here evaluated at equal times and quantifying how much the diffusive limit set by $D_{_1}$ is displaced by the slower chain molecule.
To dismiss negative values, a time-ordering operation may be carried out by introducing, for any $W_k$, the time $T_{\epsilon k} = T + \epsilon / k$.
Then we benefit of the continuity of Wiener's process and take:
\be
{\sf D}^1_n = \tfrac{1}{{T}} \lim_{\epsilon \; \rightarrow \; 0^{^+}} \lim_{\substack{t' \rightarrow \; T_{\epsilon n} \\ t \; \rightarrow \;\; T_{\epsilon 1}}} \overline{(W_{_1} (t) - W_n (t'))^2} = D_{_1} - D_n
\label{eqn:19}
\ee
This term conceptually replaces $v^2$ in Lorentz-Poincar\'e's transforms.
Eq. (\ref{eqn:17}), however, denotes a spatial perturbation to the root mean square traced by the liquid molecule, and is generally different from zero.\\
To proceed, note that line elements in Eq. (\ref{eqn:16}) should be formally second-order's and, physically, still dimensioned to a length squared.
Let the increment order be defined by:
\be
\Delta^\alpha x \; \equiv \; (\Delta \sqrt[\alpha]{x})^\alpha \;\;\;\;\;\;\;\; \alpha \in \mathbb{N}, \; x \ge 0
\label{eqn:20}
\ee
one may therefore set $\alpha = 2$:
\be
(\Delta \vartheta)^2 \; - \; (\Delta \varrho_{_1})^2 \; = \; (\Delta \vartheta')^2 \; - \; (\Delta \varrho_n)^2
\label{eqn:21}
\ee
being $\vartheta^2 = D_{_1} \tau$, $\varrho^2_m = \overline{\delta^{\; 2}_m}$.
The case with $\alpha = 1$ returns however consistent results \cite{sam7} and reduces to $\alpha = 2$ upon $\Delta \tau \; \mathlarger{\sim} \; \tau$ (unless of a proportionality constant, $\tfrac{1}{4}$, on both sides of Eq. \ref{eqn:21}) \cite{*[{In symbols $\sqrt{\Delta \tau} \rightarrow \Delta \sqrt{\tau}$ see e.g. }] [{. It suffices to note $\Delta \tau \; \mathlarger{\sim} \; (\Delta \sqrt{\tau})^2$ upon $\Delta \ln \tau \; \mathlarger{\sim} \; 1$.}] sam6}.
On this basis, the BwR transforms taking the place of Lorentz-Poincar\'e's in SR read:
\begin{align}
\Delta \varrho_n & = \f{1}{\sqrt{1 - {\sf d}^1_{n}}} (\Delta \varrho_{_1} - \sqrt{{\sf d}^1_{n}} \Delta \vartheta) \nonumber
\\
\Delta \vartheta' & = \f{1}{\sqrt{1 - {\sf d}^1_{n}}} (\Delta \vartheta \; - \sqrt{{\sf d}^1_{n}} \Delta \varrho_{_1})
\label{eqn:22}
\end{align}
having set the diffusion displacement in unit of $D_{_1}$:
\be
{\sf d}^1_{n} \; = \; \f{{\sf D}^1_{n}}{D_{_1}} \; = \; 1 - \f{1}{n}
\ee
Eqs. (\ref{eqn:22}) specify a unitary matrix connecting different domains, i.e.:
\be
{\bf L}_{_1} (\Delta \rho_{_1}, \Delta \vartheta) \; = \; {\bf L}'^{-1}_n (\Delta \rho_{n}, \Delta \vartheta') \;.
\ee
As expected, they are not defined for diffusion coefficients $> D_{_1}$, ${\sf d}^1_{n} \rightarrow - {\sf d}^1_{n}$, upon which they anyway admit the trivial solution $\varrho_m = \varrho_m$ and $\vartheta' = \vartheta$.
When ${\sf d}^1_n \rightarrow 1^-$, two identities follow, expressing Einstein's law of Brownian movement in both direct and inverse transforms, i.e. $\sqrt{{\sf d}^1_{n}} \rightarrow - \; \sqrt{{\sf d}^1_{n}}$:
\be
\overline{\delta^{\; 2}_{n}} / \tau' \;\; \mathlarger{\sim} \;\; \overline{\delta^{\; 2}_{_1}} / \tau \;\; \mathlarger{\sim} \;\; D_{_1} \;.
\ee
Fig. (\ref{fig:2}) illustrates the view settled by Eq. (\ref{eqn:16}), building up a framework that formally resembles SR.
Note that Eqs. (\ref{eqn:22}) indicate a relationship between statistical domains, obviously they do not form an ordinary Lorentz-Poincar\'e-covariant coordinate transformation. 
Another point to remark is that the relative motion in BwR is the Brownian motion of the {\em displacement} from $D_{_1}$ (Eq. \ref{eqn:19}).

\section*{Universal Scaling and Gaussian Chain}

\noindent The proof of consistency of the basic spacetime scaling in the polymer picture stems naturally from the time dilation and length contraction rules entailed by Eqs. (\ref{eqn:22}).
The limit $n \gg 1$ equals to annul the additive constants (i.e. $\tau \; \mathlarger{\sim} \; \Delta \tau$), and the results are:
\begin{align}
\varrho_n & = \sqrt{\tfrac{D_{_1}}{D_n}} \; \varrho_{_1} \nonumber
\\
\tau & = \sqrt{\tfrac{D_{n}}{D_{_1}}} \; \tau'
\label{eqn:26}
\end{align}
whereas, for a system with molecularity $m = 1$, one gets again the trivial solution (${\sf d}^m_m \rightarrow 0^+$).
Eqs. (\ref{eqn:26}) can be applied to $n$ molecules of a liquid sub-ensemble, Kuhn's step size being associated to a bare molecular time scale $\tau_{_0}$.
Let $\varrho_{_1} = n l$ and $\vartheta = n \tau_{_0}$, one gets: 
\begin{align}
\varrho_n & = n l^2 \nonumber
\\
\tau & = n^2 \tau_{_0}
\label{eqn:27}
\end{align}
that is, the random walk end-to-end size and Rouse's longest relaxation time of a chain molecule in solution \cite{doi88}.
The fact that basic polymer laws may come from a BwR framework suggested further inquiries on universal scaling \cite{sam1}.\\
Finally, stepping back from BwR to SR, we suggest that Lorentz-Poincar\'e's transformations of SR admit a statistical interpretation in light of BwR and the previous experiment bringing to Eq. (\ref{eqn:4}).
We assume that proper lengths and times are expressible as sums of independent aleatory variables, and compare the results of two observers, one (${\rm K}'$) that moves with the rectilinear rod, the other (${\rm K}$) travelling at velocity equal e.g. to $- \ups$, relative to ${\rm K}'$.
The length $\ell_{_0}$ gives a natural unit of measure for the time elapsed in this measurement, and the spatial Lorentz-Poincar\'e transform becomes:
\be
{\bf X}'_n = \; \gamma_{_0} \; ({\bf X}_n - \lambda n)
\ee
where $\gamma_{_0} = 1 / \sqrt{1 - \beta^2_{_0}}$ is the Lorentz factor and $\ell_{_0} \equiv n \lambda$.
Letting correspondingly ${\bf x}'_i = \lambda {\bf z}'_i$, the quantity:
\be
\tfrac{1}{\lambda} {\bf X}'_n \; = \; \sum^n_{i = 1} {\bf z}'_i
\ee
is a sum of independent random variables, characterized by:
\be
\overline{{\bf z}'_i} \; = \; - \; \gamma_{_0} \;, \;\;\; {\rm Var} ({{\bf z}'_i}) \; = \; 1
\label{eqn:31}
\ee
A natural application of the central limit theorem \cite{pit92} now consists of partitioning time as space ($t = n \theta$) and forming the aleatory variable:
\be
{\bf L}_n \; = \; \sqrt{\tfrac{\theta}{t}} \; {\bf X}_n
\label{eqn:32}
\ee
which converges in law to a Gaussian distribution upon $n \rightarrow \infty$:
\be
{\bf L}_n \xrightarrow{\mathcal{L}} \; {\bf L} \; \sim \; {\cal N} (0, \gamma^{-2}_{_0})
\label{eqn:33}
\ee
As the unitary variance contracts to $1/\gamma^2_{_0}$, one is left with Lorentz-FitzGerald's length contraction, reintepreted in a statistical fashion.
Clearly, the inverse transform i.e.:
\be
\overline{{\bf z}'_i} \; = \; \; \gamma_{_0} \;, \;\;\; {\rm Var} ({{\bf z}'_i}) \; = \; 1
\label{eqn:34}
\ee
leaves Eq. (\ref{eqn:33}) invariant, as it should.\\
Clearly, a similar argument applies to the temporal domain as well.
As usual in deriving the time dilation rule, a rest clock is placed in ${\rm K}$ to measure a distribution of (non-dimensional) times:
\be
\tfrac{1}{\theta} \; {\bf J}_n \; = \; \sum^n_{i = 1} {\bf y}_i
\ee
still with unitary variance, and mean values:
\be
\overline{{\bf y}_i} \; = \; \pm \; \beta^2_{_0} \;, \;\;\; {\rm Var} ({{\bf y}_i}) \; = \; 1
\label{eqn:35}
\ee
In this dual case, the central limit theorem is applied to the random variable:
\be
{\bf T}'_n \; = \; \sqrt{\tfrac{\lambda}{x}} \; {\bf J}'_n
\label{eqn:36}
\ee
and, since the unit variance now dilates by $\gamma^2_{_0}$, it returns the time dilation rule:
\be
{\bf T}'_n \xrightarrow{\mathcal{L}} \; {\bf T} \; \sim \; {\cal N} (0, \gamma^2_{_0})
\label{eqn:37}
\ee
Observe that the Fourier transforms in non-dimensional time and space values (${\cal F}$) of the normal laws in Eq. (\ref{eqn:33}) ($p_{_L}$) and Eq. (\ref{eqn:37}) ($p_{_T}$) are related, e.g. symbolically:
\be
\gamma_{_0} {\cal F}^{-1} \left\{ p_{_T} \right\} \; = \; p_{_L}
\;,\;\;\;
{\cal F} \left\{ p_{_L} \right\} \; = \; \gamma_{_0} p_{_T}
\ee
and, interestingly, the following relation holds:
\be
{\rm Var} ({\bf L}) \; {\rm Var} ({\bf T}) \; = \; 1
\ee

\section*{Acknowledgement}

The writer takes this opportunity to thank Ran Friedman (Kalmar University, Sweden) and LINXS (Lund University, Sweden) for kind hospitality and financial support.
Former discussions with Marek Grzelczak (CSIC/DIPC, Spain) are also acknowledged.

\bibliographystyle{apsrev4-2}

\bibliography{apssamp}

\providecommand{\noopsort}[1]{}\providecommand{\singleletter}[1]{#1}
\begin{thebibliography}{26}%
\makeatletter
\providecommand \@ifxundefined [1]{%
 \@ifx{#1\undefined}
}%
\providecommand \@ifnum [1]{%
 \ifnum #1\expandafter \@firstoftwo
 \else \expandafter \@secondoftwo
 \fi
}%
\providecommand \@ifx [1]{%
 \ifx #1\expandafter \@firstoftwo
 \else \expandafter \@secondoftwo
 \fi
}%
\providecommand \natexlab [1]{#1}%
\providecommand \enquote  [1]{``#1''}%
\providecommand \bibnamefont  [1]{#1}%
\providecommand \bibfnamefont [1]{#1}%
\providecommand \citenamefont [1]{#1}%
\providecommand \href@noop [0]{\@secondoftwo}%
\providecommand \href [0]{\begingroup \@sanitize@url \@href}%
\providecommand \@href[1]{\@@startlink{#1}\@@href}%
\providecommand \@@href[1]{\endgroup#1\@@endlink}%
\providecommand \@sanitize@url [0]{\catcode `\\12\catcode `\$12\catcode
  `\&12\catcode `\#12\catcode `\^12\catcode `\_12\catcode `\%12\relax}%
\providecommand \@@startlink[1]{}%
\providecommand \@@endlink[0]{}%
\providecommand \url  [0]{\begingroup\@sanitize@url \@url }%
\providecommand \@url [1]{\endgroup\@href {#1}{\urlprefix }}%
\providecommand \urlprefix  [0]{URL }%
\providecommand \Eprint [0]{\href }%
\providecommand \doibase [0]{https://doi.org/}%
\providecommand \selectlanguage [0]{\@gobble}%
\providecommand \bibinfo  [0]{\@secondoftwo}%
\providecommand \bibfield  [0]{\@secondoftwo}%
\providecommand \translation [1]{[#1]}%
\providecommand \BibitemOpen [0]{}%
\providecommand \bibitemStop [0]{}%
\providecommand \bibitemNoStop [0]{.\EOS\space}%
\providecommand \EOS [0]{\spacefactor3000\relax}%
\providecommand \BibitemShut  [1]{\csname bibitem#1\endcsname}%
\let\auto@bib@innerbib\@empty
\bibitem [{\citenamefont {Mezzasalma}(2008)}]{sam1}%
  \BibitemOpen
  \bibfield  {author} {\bibinfo {author} {\bibfnamefont {S.~A.}\ \bibnamefont
  {Mezzasalma}},\ }\href
  {https://doi.org/https://doi.org/10.1016/S1573-4285(07)00009-9} {\emph
  {\bibinfo {title} {Macromolecules in Solution and Brownian Relativity}}}\
  (\bibinfo  {publisher} {Academic Press-Elsevier, London},\ \bibinfo {year}
  {2008})\BibitemShut {NoStop}%
\bibitem [{\citenamefont {Dudley}(1966)}]{dud66}%
  \BibitemOpen
  \bibfield  {author} {\bibinfo {author} {\bibfnamefont {R.~M.}\ \bibnamefont
  {Dudley}},\ }\href {https://doi.org/10.1007/BF02592032} {\bibfield  {journal}
  {\bibinfo  {journal} {Ark. Mat.}\ }\textbf {\bibinfo {volume} {6}},\ \bibinfo
  {pages} {241} (\bibinfo {year} {1966})}\BibitemShut {NoStop}%
\bibitem [{\citenamefont {Hakim}(1968)}]{hak68}%
  \BibitemOpen
  \bibfield  {author} {\bibinfo {author} {\bibfnamefont {R.}~\bibnamefont
  {Hakim}},\ }\href {https://doi.org/10.1063/1.1664513} {\bibfield  {journal}
  {\bibinfo  {journal} {J. Math. Phys.}\ }\textbf {\bibinfo {volume} {9}},\
  \bibinfo {pages} {1805} (\bibinfo {year} {1968})},\ \Eprint
  {https://arxiv.org/abs/https://doi.org/10.1063/1.1664513}
  {https://doi.org/10.1063/1.1664513} \BibitemShut {NoStop}%
\bibitem [{\citenamefont {Dunkel}\ and\ \citenamefont {Hänggi}(2009)}]{dun09}%
  \BibitemOpen
  \bibfield  {author} {\bibinfo {author} {\bibfnamefont {J.}~\bibnamefont
  {Dunkel}}\ and\ \bibinfo {author} {\bibfnamefont {P.}~\bibnamefont
  {Hänggi}},\ }\href
  {https://doi.org/https://doi.org/10.1016/j.physrep.2008.12.001} {\bibfield
  {journal} {\bibinfo  {journal} {Phys. Rep.}\ }\textbf {\bibinfo {volume}
  {471}},\ \bibinfo {pages} {1 } (\bibinfo {year} {2009})}\BibitemShut
  {NoStop}%
\bibitem [{\citenamefont {de~Gennes}(1979)}]{pierre79}%
  \BibitemOpen
  \bibfield  {author} {\bibinfo {author} {\bibfnamefont {P.}~\bibnamefont
  {de~Gennes}},\ }\href@noop {} {\emph {\bibinfo {title} {Scaling Concepts in
  Polymer Physics}}}\ (\bibinfo  {publisher} {Cornell University Press},\
  \bibinfo {address} {Ithaca},\ \bibinfo {year} {1979})\BibitemShut {NoStop}%
\bibitem [{\citenamefont {Nienhuis}(1982)}]{nie82}%
  \BibitemOpen
  \bibfield  {author} {\bibinfo {author} {\bibfnamefont {B.}~\bibnamefont
  {Nienhuis}},\ }\href {https://doi.org/10.1103/PhysRevLett.49.1062} {\bibfield
   {journal} {\bibinfo  {journal} {Phys. Rev. Lett.}\ }\textbf {\bibinfo
  {volume} {49}},\ \bibinfo {pages} {1062} (\bibinfo {year}
  {1982})}\BibitemShut {NoStop}%
\bibitem [{\citenamefont {Rubinstein}\ and\ \citenamefont
  {Colby}(2003)}]{rub03}%
  \BibitemOpen
  \bibfield  {author} {\bibinfo {author} {\bibfnamefont {M.}~\bibnamefont
  {Rubinstein}}\ and\ \bibinfo {author} {\bibfnamefont {R.}~\bibnamefont
  {Colby}},\ }\href@noop {} {\emph {\bibinfo {title} {Polymer Physics}}}\
  (\bibinfo  {publisher} {Oxford University Press},\ \bibinfo {year} {2003})\
  p.\ \bibinfo {pages} {440}\BibitemShut {NoStop}%
\bibitem [{\citenamefont {Mezzasalma}(2007{\natexlab{a}})}]{sam2}%
  \BibitemOpen
  \bibfield  {author} {\bibinfo {author} {\bibfnamefont {S.~A.}\ \bibnamefont
  {Mezzasalma}},\ }\href {https://doi.org/10.1021/la701891m} {\bibfield
  {journal} {\bibinfo  {journal} {Langmuir}\ }\textbf {\bibinfo {volume}
  {23}},\ \bibinfo {pages} {12737} (\bibinfo {year}
  {2007}{\natexlab{a}})}\BibitemShut {NoStop}%
\bibitem [{\citenamefont {Mezzasalma}(2006)}]{sam4}%
  \BibitemOpen
  \bibfield  {author} {\bibinfo {author} {\bibfnamefont {S.~A.}\ \bibnamefont
  {Mezzasalma}},\ }\href {https://doi.org/10.1021/jp063539d} {\bibfield
  {journal} {\bibinfo  {journal} {J. Phys. Chem. B}\ }\textbf {\bibinfo
  {volume} {110}},\ \bibinfo {pages} {23507} (\bibinfo {year}
  {2006})}\BibitemShut {NoStop}%
\bibitem [{\citenamefont {Mezzasalma}(2007{\natexlab{b}})}]{sam5}%
  \BibitemOpen
  \bibfield  {author} {\bibinfo {author} {\bibfnamefont {S.~A.}\ \bibnamefont
  {Mezzasalma}},\ }\href
  {https://doi.org/https://doi.org/10.1016/j.chemphys.2007.03.008} {\bibfield
  {journal} {\bibinfo  {journal} {Chem. Phys.}\ }\textbf {\bibinfo {volume}
  {334}},\ \bibinfo {pages} {232 } (\bibinfo {year}
  {2007}{\natexlab{b}})}\BibitemShut {NoStop}%
\bibitem [{\citenamefont {Brouzet}\ \emph {et~al.}(2014)\citenamefont
  {Brouzet}, \citenamefont {Verhille},\ and\ \citenamefont {Le~Gal}}]{bro14}%
  \BibitemOpen
  \bibfield  {author} {\bibinfo {author} {\bibfnamefont {C.}~\bibnamefont
  {Brouzet}}, \bibinfo {author} {\bibfnamefont {G.}~\bibnamefont {Verhille}},\
  and\ \bibinfo {author} {\bibfnamefont {P.}~\bibnamefont {Le~Gal}},\ }\href
  {https://doi.org/10.1103/PhysRevLett.112.074501} {\bibfield  {journal}
  {\bibinfo  {journal} {Phys. Rev. Lett.}\ }\textbf {\bibinfo {volume} {112}},\
  \bibinfo {pages} {074501} (\bibinfo {year} {2014})}\BibitemShut {NoStop}%
\bibitem [{\citenamefont {Verhille}\ and\ \citenamefont
  {Bartoli}(2016)}]{ver16}%
  \BibitemOpen
  \bibfield  {author} {\bibinfo {author} {\bibfnamefont {G.}~\bibnamefont
  {Verhille}}\ and\ \bibinfo {author} {\bibfnamefont {A.}~\bibnamefont
  {Bartoli}},\ }\href {https://doi.org/10.1007/s00348-016-2201-1} {\bibfield
  {journal} {\bibinfo  {journal} {Exp. Fluids}\ }\textbf {\bibinfo {volume}
  {57}},\ \bibinfo {pages} {117} (\bibinfo {year} {2016})}\BibitemShut
  {NoStop}%
\bibitem [{\citenamefont {Odijk}(2017)}]{theo17}%
  \BibitemOpen
  \bibfield  {author} {\bibinfo {author} {\bibfnamefont {T.}~\bibnamefont
  {Odijk}},\ }\href@noop {} {\bibfield  {journal} {\bibinfo  {journal} {Physica
  A}\ }\textbf {\bibinfo {volume} {467}},\ \bibinfo {pages} {180} (\bibinfo
  {year} {2017})}\BibitemShut {NoStop}%
\bibitem [{\citenamefont {Le~Nov\'ere}(2015)}]{nov15}%
  \BibitemOpen
  \bibfield  {author} {\bibinfo {author} {\bibfnamefont {N.}~\bibnamefont
  {Le~Nov\'ere}},\ }\href {https://doi.org/10.1038/nrg3885} {\bibfield
  {journal} {\bibinfo  {journal} {Nat. Rev. Genet.}\ }\textbf {\bibinfo
  {volume} {16}},\ \bibinfo {pages} {146 } (\bibinfo {year}
  {2015})}\BibitemShut {NoStop}%
\bibitem [{\citenamefont {Cowan}\ \emph {et~al.}(2012)\citenamefont {Cowan},
  \citenamefont {Moraru}, \citenamefont {Schaff}, \citenamefont {Slepchenko},\
  and\ \citenamefont {Loew}}]{cow12}%
  \BibitemOpen
  \bibfield  {author} {\bibinfo {author} {\bibfnamefont {A.~E.}\ \bibnamefont
  {Cowan}}, \bibinfo {author} {\bibfnamefont {I.~I.}\ \bibnamefont {Moraru}},
  \bibinfo {author} {\bibfnamefont {J.~C.}\ \bibnamefont {Schaff}}, \bibinfo
  {author} {\bibfnamefont {B.~M.}\ \bibnamefont {Slepchenko}},\ and\ \bibinfo
  {author} {\bibfnamefont {L.~M.}\ \bibnamefont {Loew}},\ }in\ \href
  {https://doi.org/https://doi.org/10.1016/B978-0-12-388403-9.00008-4} {\emph
  {\bibinfo {booktitle} {Computational Methods in Cell Biology}}},\ Vol.\
  \bibinfo {volume} {110}\ (\bibinfo  {publisher} {Academic Press},\ \bibinfo
  {year} {2012})\ pp.\ \bibinfo {pages} {195 -- 221}\BibitemShut {NoStop}%
\bibitem [{\citenamefont {Gillespie}(2007)}]{gil07}%
  \BibitemOpen
  \bibfield  {author} {\bibinfo {author} {\bibfnamefont {D.~T.}\ \bibnamefont
  {Gillespie}},\ }\href
  {https://doi.org/10.1146/annurev.physchem.58.032806.104637} {\bibfield
  {journal} {\bibinfo  {journal} {Annu. Rev. Phys. Chem.}\ }\textbf {\bibinfo
  {volume} {58}},\ \bibinfo {pages} {35} (\bibinfo {year} {2007})}\BibitemShut
  {NoStop}%
\bibitem [{\citenamefont {Isaacson}(2009)}]{isaac09}%
  \BibitemOpen
  \bibfield  {author} {\bibinfo {author} {\bibfnamefont {S.~A.}\ \bibnamefont
  {Isaacson}},\ }\href {https://doi.org/10.1137/070705039} {\bibfield
  {journal} {\bibinfo  {journal} {SIAM J. Appl. Math.}\ }\textbf {\bibinfo
  {volume} {70}},\ \bibinfo {pages} {77} (\bibinfo {year} {2009})}\BibitemShut
  {NoStop}%
\bibitem [{\citenamefont {Misner}\ \emph {et~al.}(2017)\citenamefont {Misner},
  \citenamefont {Thorne}, \citenamefont {Wheeler},\ and\ \citenamefont
  {Kaiser}}]{mis17}%
  \BibitemOpen
  \bibfield  {author} {\bibinfo {author} {\bibfnamefont {C.}~\bibnamefont
  {Misner}}, \bibinfo {author} {\bibfnamefont {K.}~\bibnamefont {Thorne}},
  \bibinfo {author} {\bibfnamefont {J.}~\bibnamefont {Wheeler}},\ and\ \bibinfo
  {author} {\bibfnamefont {D.}~\bibnamefont {Kaiser}},\ }\href@noop {} {\emph
  {\bibinfo {title} {Gravitation}}}\ (\bibinfo  {publisher} {Princeton
  University Press},\ \bibinfo {year} {2017})\BibitemShut {NoStop}%
\bibitem [{\citenamefont {It\^{o}}(1951)}]{ito51}%
  \BibitemOpen
  \bibfield  {author} {\bibinfo {author} {\bibfnamefont {K.}~\bibnamefont
  {It\^{o}}},\ }\href@noop {} {\bibfield  {journal} {\bibinfo  {journal} {Mem.
  Amer. Math. Soc.}\ }\textbf {\bibinfo {volume} {4}},\ \bibinfo {pages} {51}
  (\bibinfo {year} {1951})}\BibitemShut {NoStop}%
\bibitem [{\citenamefont {Nelson}(1967)}]{nel67}%
  \BibitemOpen
  \bibfield  {author} {\bibinfo {author} {\bibfnamefont {E.}~\bibnamefont
  {Nelson}},\ }\href {https://books.google.it/books?id=xuQ9DwAAQBAJ} {\emph
  {\bibinfo {title} {Dynamical Theories of Brownian Motion}}},\ Mathematical
  Notes\ (\bibinfo  {publisher} {Princeton University Press},\ \bibinfo {year}
  {1967})\BibitemShut {NoStop}%
\bibitem [{\citenamefont {Van~Kampen}(2007)}]{van07}%
  \BibitemOpen
  \bibfield  {author} {\bibinfo {author} {\bibfnamefont {N.~G.}\ \bibnamefont
  {Van~Kampen}},\ }\href@noop {} {\emph {\bibinfo {title} {Stochastic Processes
  in Physics and Chemistry}}}\ (\bibinfo  {publisher} {North Holland},\
  \bibinfo {address} {Amsterdam},\ \bibinfo {year} {2007})\BibitemShut
  {NoStop}%
\bibitem [{\citenamefont {Brze{\'z}niak}\ and\ \citenamefont
  {Zastawniak}(2000)}]{brz00}%
  \BibitemOpen
  \bibfield  {author} {\bibinfo {author} {\bibfnamefont {Z.}~\bibnamefont
  {Brze{\'z}niak}}\ and\ \bibinfo {author} {\bibfnamefont {T.}~\bibnamefont
  {Zastawniak}},\ }\href@noop {} {\emph {\bibinfo {title} {Basic Stochastic
  Processes}}}\ (\bibinfo  {publisher} {Springer-Verlag},\ \bibinfo {address}
  {London},\ \bibinfo {year} {2000})\BibitemShut {NoStop}%
\bibitem [{\citenamefont {Mezzasalma}(2005{\natexlab{a}})}]{sam7}%
  \BibitemOpen
  \bibfield  {author} {\bibinfo {author} {\bibfnamefont {S.~A.}\ \bibnamefont
  {Mezzasalma}},\ }\href
  {https://doi.org/https://doi.org/10.1016/j.cplett.2005.01.027} {\bibfield
  {journal} {\bibinfo  {journal} {Chem. Phys. Lett.}\ }\textbf {\bibinfo
  {volume} {403}},\ \bibinfo {pages} {334 } (\bibinfo {year}
  {2005}{\natexlab{a}})}\BibitemShut {NoStop}%
\bibitem [{\citenamefont {Mezzasalma}(2005{\natexlab{b}})}]{sam6}%
  \BibitemOpen
  \bibfield  {author} {\bibinfo {author} {\bibfnamefont {S.~A.}\ \bibnamefont
  {Mezzasalma}},\ }\href {https://doi.org/10.1016/j.aop.2005.01.007} {\bibfield
   {journal} {\bibinfo  {journal} {Ann. Phys.}\ }\textbf {\bibinfo {volume}
  {318}},\ \bibinfo {pages} {408} (\bibinfo {year}
  {2005}{\natexlab{b}})}\BibitemShut {NoStop}%
\bibitem [{\citenamefont {Doi}\ and\ \citenamefont {Edwards}(1988)}]{doi88}%
  \BibitemOpen
  \bibfield  {author} {\bibinfo {author} {\bibfnamefont {M.}~\bibnamefont
  {Doi}}\ and\ \bibinfo {author} {\bibfnamefont {S.}~\bibnamefont {Edwards}},\
  }\href@noop {} {\emph {\bibinfo {title} {The Theory of Polymer Dynamics}}}\
  (\bibinfo  {publisher} {Clarendon Press, Oxford},\ \bibinfo {year}
  {1988})\BibitemShut {NoStop}%
\bibitem [{\citenamefont {Pitman}(1992)}]{pit92}%
  \BibitemOpen
  \bibfield  {author} {\bibinfo {author} {\bibfnamefont {J.}~\bibnamefont
  {Pitman}},\ }\href@noop {} {\emph {\bibinfo {title} {Probability}}}\
  (\bibinfo  {publisher} {Springer, New York},\ \bibinfo {year}
  {1992})\BibitemShut {NoStop}%
\end{thebibliography}%

\end{document}